# Calculating Gravitational Wave Signatures from Binary Black Hole Mergers

Joan M. Centrella

*Laboratory for High Energy Astrophysics, NASA Goddard Space Flight Center, Greenbelt, MD 20771 USA*

**Abstract.** Calculations of the final merger stage of binary black hole evolution can only be carried out using full scale numerical relativity simulations. This article provides a general overview of these calculations, highlighting recent progress and current challenges.

## INTRODUCTION

The final coalescence of a binary black hole (BBH) is driven by gravitational wave emission and proceeds in 3 stages [1]. In the quasi-adiabatic *inspiral* phase, the BHs have a large enough separation that they can be treated analytically as point particles in the post-Newtonian limit. This relatively slow inspiral is followed by a dynamical *merger* phase, during which the BHs plunge toward one another and merge, forming a single BH. This highly distorted remnant then emits gravitational radiation in quasi-normal modes as it *rings down* to form a quiescent rotating BH.

Calculating the gravitational radiation signatures from BBH coalescence is essential for understanding and interpreting the data expected from both ground-based and space-based detectors [2]. Both the inspiral and ringdown stages can be handled analytically. During the inspiral, the waveform is a "chirp," which is a sinusoid increasing in both frequency and amplitude as the orbital separation shrinks. Since the detectors can typically observe thousands of orbits during inspiral, these waveforms can be used as templates for data analysis algorithms based on matched filtering to yield information on the masses and spins of the components. The final ringdown of the distorted remnant will produce a burst of gravitational radiation. The waveforms from this stage are damped sinusoids that can be calculated analytically using BH perturbation theory, and can be used in identifying the properties of the final Kerr hole.

However, the merger phase, starting at the end of the inspiral and proceeding through the plunge and dynamical merger, is governed by strong, nonlinear gravitational fields. The resulting burst waveforms can only be calculated through numerical relativity simulations of the full Einstein equations in 3 spatial dimensions (3-D) plus time. Gravitational wave observations of BH mergers can provide an outstanding probe of a regime expected to be rich in strong-field spacetime dynamics, including black hole spin flips and couplings [3].

Numerical modeling of BH mergers is a challenging enterprise, due to the deep mathematical nature of general relativity, the need to solve a fairly large number of nonlinear, coupled partial differential equations, and the wide range of physical scales involved. While we still do not have definitive waveforms starting in the late inspiral stage and proceeding through ringdown, significant progress has been made over the past few years. In this article, we provide a brief review of the status of numerical relativity BH simulations, highlighting recent successes and current challenges. While this inevitably involves some excursions into the mathematical side of general relativity, this will be done for purposes of illustration rather than rigor. Further details and additional references for interested readers can be found in the recent review by Baumgarte and Shapiro [4].

## EINSTEIN'S EQUATIONS IN 3+1 FORM

Provided that any accompanying accretion disks are dynamically negligible, BBHs coalescing through the emission of gravitational radiation are solutions to the vacuum (i.e., source-free) Einstein equations of general relativity. As such, their timescales are proportional to the total mass $M$ and, after time scaling, all other properties of the dynamics and waveforms depend only on ratios involving the masses and spins of the components. This has the pleasant consequence that calculations of the gravitational waveforms from all three stages of BBH coalescence can be easily scaled to apply to any masses and spins [5].

The Einstein equations in vacuum are [6]

$$^{(4)}R_{\mu\nu} - \tfrac{1}{2}{}^{(4)}Rg_{\mu\nu} = 0, \tag{1}$$

where we use spacetime indices $\mu, \nu = 0, 1, 2, 3$. Here, $^{(4)}R_{\mu\nu}$ is the 4-D or spacetime Ricci tensor, $^{(4)}R \equiv {}^{(4)}R_{\mu\nu}g^{\mu\nu}$, and we assume an implied summation over pairs of repeated up and down indices. Since all second rank tensors used here are symmetric (e.g., $^{(4)}R_{\mu\nu} = {}^{(4)}R_{\nu\mu}$), Eqs. (1) comprise 10 coupled, nonlinear partial differential equations for the evolution of the spacetime metric $g_{\mu\nu}$ which are second order in both time and space derivatives.

To evolve BBH mergers using numerical relativity, we must first write these equations in a form suitable for solution on a computer. We will accomplish this using the so-called 3 + 1 spacetime split, in which 4-D spacetime is envisioned as being sliced into a collection of 3-D spacelike hypersurfaces, threaded by a congruence of timelike curves along which time $t$ is measured [6].

The coordinate freedom of general relativity provides 4 freely specifiable coordinate conditions or gauge choices. In the 3 + 1 framework, these are the *lapse function* $\alpha$, which measures the lapse of proper time $\alpha\delta t$ between neighboring slices, and the 3 components of the shift vector $\beta^i$, which governs the movement of spatial coordinates during evolution from slice to slice. The spacetime metric takes the form

$$g_{\mu\nu} = \begin{pmatrix} -\alpha^2 + \beta_k\beta^k & \beta_i \\ \beta_j & \gamma_{ij} \end{pmatrix}, \tag{2}$$

where $\gamma_{ij}$ is the spatial metric of the 3-D slice, $\beta_i = \gamma^{ik}\beta_k$, and spatial indices $i, j = 1, 2, 3$. We use units in which $c = G = 1$. The choices made for $\alpha$ and $\beta^i$ are critical for successful BH evolutions, as discussed below.

The Einstein equations (1) split into 2 sets. Four equations set conditions on the spacelike slices: the *Hamiltonian constraint*

$$R + K^2 - K_{ij}K^{ij} = 0 \tag{3}$$

and the *momentum constraints*

$$D_j K^j{}_i - D_i K = 0. \tag{4}$$

Here, $R \equiv \gamma^{ij} R_{ij}$ is the Ricci scalar and $D_j$ is the spatial covariant derivative compatible with the 3-metric $\gamma_{ij}$ in a spacelike slice. $K_{ij}$ is the extrinsic curvature of the slice, and $K \equiv K^i{}_i = \gamma^{ij} K_{ij}$ is its trace. The constraint equations (3) and (4) also set conditions on the initial data, as discussed in the next section. The remaining 6 equations give the evolution of the initial data in time. Taking $\gamma_{ij}$ and $K_{ij}$ to be the basic variables, we can write them as a set of equations that are first order in time. In the standard Arnowit-Deser-Misner (ADM) [7, 4] formalism, these become the evolution equations for the 3-metric $\gamma_{ij}$,

$$\partial_t \gamma_{ij} = -2\alpha K_{ij} + D_i \beta_j + D_j \beta_i, \tag{5}$$

and for the extrinsic curvature $K_{ij}$,

$$\partial_t K_{ij} = -D_i D_j \alpha + \alpha(R_{ij} - 2K_{ik}K^k{}_j + KK_{ij}) + \beta^k D_k K_{ij} + K_{ik} D_j \beta^k + K_{kj} D_i \beta^k. \tag{6}$$

Note that the terms involving $R_{ij}$ and $R$ contain both first and second spatial derivatives of $\gamma_{ij}$.

## INITIAL DATA FOR BBH MERGERS

Numerical simulations of BBH mergers should begin with initial conditions characterizing the late stages of the inspiral, before the plunge begins. Since the binaries reach this state due to the emission of gravitational radiation, realistic initial conditions must take account of this radiation as well as the positions and momenta of the BHs. Techniques are available for producing BBH initial data that satisfies the Einstein equations, and current efforts are focused on insuring that the data represents the astrophysical state of the binary accurately.

Setting up initial data for a numerical relativity simulation requires solving the 4 constraint equations (3) and (4) on a spacelike slice for $\gamma_{ij}$ and $K_{ij}$ [8]. Most efforts to solve these equations begin with a conformal decomposition of the metric and extrinsic curvature. Specifically, the physical 3-metric $\gamma_{ij}$ is written

$$\gamma_{ij} = \psi^4 \tilde{\gamma}_{ij}, \tag{7}$$

where $\psi$ is known as the conformal factor and $\tilde{\gamma}_{ij}$ is the conformal metric. The extrinsic curvature is split into its trace $K$ and a trace-free part $A_{ij}$, so that

$$K_{ij} = A_{ij} + \tfrac{1}{3}\gamma_{ij}K. \tag{8}$$

Various approaches then differ by the way in which $A_{ij}$ is decomposed [9].

In all of these techniques, the initial data variables separate into 2 sets: those variables that are freely specifiable (such as the conformal metric $\tilde{\gamma}_{ij}$, the trace $K$, and some parts of $A_{ij}$), and those that are constrained (typically, the conformal factor $\psi$ and other parts of $A_{ij}$). Once the freely specifiable quantities are chosen, they are inserted into the conformally transformed initial value equations; these nonlinear elliptic equations are then solved for the constrained parts of the data. Combining the free and constrained pieces then produces a valid set $(\gamma_{ij}, K_{ij})$ of initial data that obeys the Einstein initial value equations (3) and (4).

In most of the work on the initial value problem to date, the freely specifiable data have been chosen to decouple and otherwise simplify the solution of the constraint equations. For example, we can choose the initial slice to be conformally flat ($\tilde{\gamma}_{ij} = f_{ij}$, the metric of flat space) and have $K = 0$ (known as maximal slicing). With a particular decomposition of $A_{ij}$, this not only decouples the constraint equations but also allows analytic solutions of the momentum constraints for a BH with both linear momentum and spin [10]. Since the momentum constraints become linear equations for the constrained parts of $A_{ij}$ in this approach, we can superpose these solutions to obtain data for any number of BHs with linear and angular momentum. The nonlinear Hamiltonian constraint must still be solved numerically. This type of data, pioneered by Bowen and York [10] has formed the starting point for various numerical simulations of BHs, including grazing collisions [11] and merger calculations beginning near the final plunge [12, 13].

Much current work centers on producing BBH initial data sets that are astrophysically realistic. In general, this involves choosing spatial metrics that are not conformally flat. Some recent work has attempted to incorporate the results of post-Newtonian expansions near the end of the inspiral phase into the freely specifiable data (e.g., [14]) while other efforts have used a non-flat conformal 3-metric based on superposing 2 BHs in Kerr-Schild coordinates (e.g., [15]). A special challenge arises because, in the conformal decomposition process, the physical metric $\gamma_{ij}$ and the extrinsic curvature $K_{ij}$ that emerge out of the solution process can be changed significantly from the freely specifiable quantities that are chosen at the start. The conformal thin-sandwich decomposition [16], which includes information on the evolution of the metric away from the initial slice, may alleviate such problems. It is also a promising approach for constructing BBH initial data on quasi-circular orbits near the plunge [4].

All of these techniques can introduce unphysical gravitational waves into the initial data. For example, a recent study constructed BBH initial data sets using similar choices for the freely specifiable data but different conformal decompositions [9]. The amount of gravitational radiation present differed among the resulting initial data sets by a few percent of the total mass, an amount that is comparable with the total gravitational radiation expected from the entire merger process.

# EVOLVING BBH MERGERS

The BBH initial data must be evolved for a number of orbits prior to the plunge at the end of the inspiral, then through the plunge and merger, and into the ringdown stage [17]. For a BBH system near the plunge with separation $a \sim 6M$, the orbital period $P \sim 90M$. We therefore need to be able to evolve the system for $\sim 1000M$ or longer, to produce astrophysically relevant waveforms useful for detection.

Over the past decade, considerable effort has been expended by the worldwide numerical relativity community towards achieving this goal. While significant progress has been made, a number of difficult challenges remain. In this section we'll present a brief overview of BBH evolutions, highlighting both current successes and outstanding problems.

## Choice of Formalism

The ADM evolution equations (5) and (6) mark the starting point for most BH simulations. However, a direct numerical implementation of these equations in 3-D quickly develops exponentially growing unstable modes that cause the code to crash. In particular, a BBH evolution using the standard ADM equations lasts only $\sim 13M$ [12], a small fraction of an orbital period near the start of the plunge. While the exact cause of these instabilities is not yet understood, we do know that they are properties of the mathematical formulation of the Einstein equations, and not of the numerical methods employed [18].

Baumgarte and Shapiro [19] and Shibata and Nakamura [20] modified the original ADM evolution equations (5) and (6) using a conformal-traceless decomposition. This so-called BSSN formalism is obtained by writing the conformal factor in the form $\psi = e^\phi$, so that

$$\tilde{\gamma}_{ij} = e^{-4\phi} \gamma_{ij}, \tag{9}$$

with the choice $\tilde{\gamma} \equiv \det(\tilde{\gamma}_{ij}) = 1$. The traceless part of $K_{ij}$ is scaled according to

$$\tilde{A}_{ij} = e^{-4\phi} A_{ij}. \tag{10}$$

The conformal connection functions

$$\tilde{\Gamma}^i \equiv \tilde{\gamma}^{jk} \tilde{\Gamma}^i_{jk} = -\partial_j \tilde{\gamma}^{ij}, \tag{11}$$

are introduced as new independent variables. Here, the $\bar{\Gamma}^i_{jk}$ are the connection coefficients (or Christoffel symbols) associated with $\tilde{\gamma}_{ij}$, and the second equality in (11) relies on the condition $\tilde{\gamma} = 1$. The ADM evolution equations then take the following form:

$$\partial_t \phi = -\tfrac{1}{6}\alpha K + \beta^i \partial_i \phi + \tfrac{1}{6}\partial_i \beta^i, \tag{12}$$

$$\partial_t \tilde{\gamma}_{ij} = -2\alpha \tilde{A}_{ij} + \beta^k \partial_k \tilde{\gamma}_{ij} + \tilde{\gamma}_{ik}\partial_j \beta^k + \tilde{\gamma}_{kj}\partial_i \beta^k - \tfrac{2}{3}\tilde{\gamma}_{ij}\partial_k \beta^k, \tag{13}$$

$$\partial_t K = -\gamma^{ij} D_j D_i \alpha + \alpha(\tilde{A}_{ij}\tilde{A}^{ij} + \tfrac{1}{3}K^2) + \beta^i \partial_i K, \tag{14}$$

$$\partial_t \tilde{A}_{ij} = e^{-4\phi}\left(-(D_i D_j \alpha)^{TF} + \alpha(R_{ij}^{TF})\right) + \alpha(K\tilde{A}_{ij} - 2\tilde{A}_{il}\tilde{A}^l{}_j) \\ + \beta^k \partial_k \tilde{A}_{ij} + \tilde{A}_{ik}\partial_j \beta^k + \tilde{A}_{kj}\partial_i \beta^k - \tfrac{2}{3}\tilde{A}_{ij}\partial_k \beta^k, \quad (15)$$

$$\partial_t \tilde{\Gamma}^i = -2\tilde{A}^{ij}\partial_j \alpha + 2\alpha\left(\tilde{\Gamma}^i_{jk}\tilde{A}^{kj} - \tfrac{2}{3}\tilde{\gamma}^{ij}\partial_j K + 6\tilde{A}^{ij}\partial_j \phi\right) \\ + \beta^j \partial_j \tilde{\Gamma}^i - \tilde{\Gamma}^j \partial_j \beta^i + \tfrac{2}{3}\tilde{\Gamma}^i \partial_j \beta^j + \tfrac{1}{3}\tilde{\gamma}^{li}\partial_l \partial_i \beta^j + \tilde{\gamma}^{lj}\partial_l \partial_j \beta^i, \quad (16)$$

where the superscript TF in (15) indicates the trace-free part (e.g. $R_{ij}^{TF} = R_{ij} - \gamma_{ij}R/3$) and the definition (11) serves as a new constraint equation. Using $\tilde{\Gamma}^i$ and its first spatial derivatives, the Ricci tensor $R_{ij}$ is then written so that the only second spatial derivatives of $\tilde{\gamma}_{ij}$ that remain are in the wave operator $\tilde{\gamma}^{lm}\partial_l \partial_m \tilde{\gamma}_{ij}$; see [19, 4] for details.

The BSSN equations (12) - (16) constitute a set of 17 evolution equations. In a typical simulation, all of these equations are updated, and the conditions $\tilde{\gamma} = 1$ and $\text{trace}(A_{ij}) \equiv A^i{}_j = 0$ are then imposed. One or more of the new constraints (11) may also be imposed.

Numerical implementations of the original BSSN formalism given here exhibit much better stability properties than those using the standard ADM equations [21]. Further improvements and variations have been introduced, mostly based on adding some of the constraint equations to the evolution equations, or enforcing some of the constraints during evolution [22, 23] (see also [24]). With these developments, single rotating BHs can now be evolved stably for several thousand $M$ [23] and BBHs for $> 100M$ [25].

A somewhat different approach to developing improved formalisms centers on writing the Einstein equations in hyperbolic form. Current efforts along these lines have led to the discovery of many-paramenter families of representations of the Einstein equations. For a given set of parameters, one obtains a relatively large set of equations that are fully first order in both time and space derivatives [26, 27]. The stability of these systems is strongly dependent on which set of parameters, and hence which representation of the Einstein equations, is chosen [18, 26]. Currently, evolutions of a single BH are possible for up to several thousand $M$ for optimal parameter choices [28]. Although these systems are mathematically elegant, the large numbers of equations involved can make them unattractive from the point of view of numerical implementation. However, they do have the advantage that their characteristic structure can be analyzed readily and used to impose natural boundary conditions both at the outer edges of a numerical grid and on inner excision boundaries of BHs [26].

## Representing a BH on a Numerical Grid

A BH contains a physical singularity, at which the curvature is infinite. Numerical relativity simulations involving BHs thus face special challenges in representing these objects on a computational grid. Several techniques have been developed to accomplish this task.

The spatial metric for a Schwarzschild BH in isotropic coordinates is conformally flat, so that $\gamma_{ij} = \psi^4 f_{ij}$. Taking $K_{ij} = 0$, the momentum constraints are trivially satisfied and the Hamiltonian constraint becomes $\Delta^{\text{flat}}\psi = 0$, where $\Delta^{\text{flat}}$ is the Laplacian in flat space. With the asymptotically flat boundary condition $\psi \to 1$ as $r \to \infty$, the conformal factor takes the expected form $\psi = 1 + M/2r$, where $M$ is the Schwarzschild mass and the event horizon is located at $r_H = M/2$. Since the Hamiltonian constraint is a linear equation in this case, we can easily add solutions to get initial data for multiple BHs. Thus the conformal factor for 2 BHs is $\psi = 1 + M_1/2r_1 + M_2/2r_2$, where the BH of mass $M_1$ is located at $(x_1, y_1, z_1)$ with $r_1^2 = (x-x_1)^2 + (y-y_1)^2 + (z-z_1)^2$, and similarly for $M_2$ [29].

For a single BH, the coordinate transformation $r' = M^2/4r$ is an isometry that maps every point inside $r_H = M/2$ into a point outside $r_H$. The geometry inside $r_H$ is thus the same as that outside $r_H$, allowing us to think of the BH as consisting of 2 asymptotically flat regions or sheets connected by a throat at $r = r_H$ [6]. For 2 BHs, we can have a single sheet with 2 throats, each of which connects to a separate asymptotically flat region, giving a 3-sheeted topology. Alternatively, we can add additional throats, corresponding to mirror images of the companion BH, inside each of the original throats. This conformal imaging approach restores the isometry and yields a 2-sheeted topology in which the throats connect identical asymptotically flat sheets [30]. To implement this in numerical relativity, we can use the isometry conditions on the throats as boundary conditions.

A similar procedure can be carried out following the Bowen-York [10] approach to yield data for BHs with spins and linear momenta. As discussed above, we assume conformal flatness and take $K = 0$. Then $A_{ij}$ takes a specific form that solves the momentum constraints; since these are linear equations in this case, the solutions can be added together. The Hamiltonian constraint is a nonlinear equation that must be solved numerically. Isometry boundary conditions have to be applied at the throats, which are rather complicated surfaces [31].

The puncture method eliminates the need to apply isometry conditions on the throats [32]. We consider the initial slice to have points, called punctures, removed at $r_1$ and $r_2$. The conformal factor is written as the sum $\psi = u + \psi'$, where $\psi' = 1 + \mathcal{M}_1/2r_1 + \mathcal{M}_2/2r_2$, and the constants $\mathcal{M}_1 \to M_1$ and $\mathcal{M}_2 \to M_2$ as the separation between the BHs approaches infinity. The Hamiltonian constraint becomes a nonlinear equation for the function $u$. In this approach, the singular terms are absorbed into the analytic term $\psi'$ and the remaining function $u$ is regular everywhere, even at the punctures $r_1$ and $r_2$. The puncture technique has also been extended to post-Newtonian data [14].

Puncture data can be evolved by choosing the location of the punctures so that they do not coincide with a point on the computational grid. With suitably chosen lapse and shift, the punctures do not move through the grid (although their positions in spacetime, governed by the evolution of the metric, do change), and the spacetime in their immediate vicinity does not evolve [25]. The puncture method has been used to evolve single and multiple BH systems, including a grazing collision [11] and the final plunge [12, 13].

The other method that has been developed for handling BHs on a grid is known as excision. This approach relies on the fact that no physical information can progagate from inside the BH event horizon to influence the spacetime outside. Thus, it is not necessary to evolve the Einstein equations on the spacetime inside the horizon, and the interior of the BH can be excised from the numerical grid [33]

The event horizon is the boundary between those spacetime events which emit light rays reaching infinity, and those which do not. Since the event horizon is a global quantity, it is necessary to know the entire spacetime before the location of the event horizon can be found. However, this is generally not possible during a numerical simulation, in which the spacetime is constructed by evolving from one slice to the next. In practice, a related concept known as an apparent horizon is used. The location of the apparent horizon can be calculated on each spacelike slice using $\gamma_{ij}$ and $K_{ij}$. Since the apparent horizon is always located inside the event horizon [34], it is safe to excise zones within the apparent horizon. In the case of Schwarzschild and Kerr BHs, the locations of the apparent and event horizons coincide.

In a typical implementation of excision, several buffer zones separate the apparent horizon from the excised region inside. Several different shapes have been used to define the excision boundary, including a cube and a sphere. An appropriate discretization of the Einstein equations must be used for the grid points on the excision boundary. And, when an excised BH is moved, points that were previously within the excised region become points on the computational grid. Special extrapolation techniques are used to populate these points with physical data [35].

Excision has been used to evolve both static and distorted single BHs [23, 36], single BHs that move across a grid [37, 35], and a grazing collision of 2 BHs [38].

## Choice of Lapse and Shift

A key ingredient of successful numerical relativity simulations is the choice of suitable coordinate conditions. Within the 3 + 1 framework, these are specified by the lapse function $\alpha$, which controls the slicing of spacetime, and the shift vector $\beta^i$, which allows the spatial coordinates to move [39].

The simplest choice is geodesic slicing, in which $\alpha = 1$ and $\beta^i = 0$. Here, the coordinates coincide with freely-falling observers that travel on geodesics normal to each spacelike slice. However, since gravity is attractive, these observers fall towards the BH and will generally hit the singularity in a finite time. This cause the code to crash before any significant dynamical evolution of the system can be accomplished.

Singularity avoiding coordinate conditions have been developed to prevent the spacelike slices from intersecting BH singularities. In this approach, conditions are imposed on the lapse function so that the resulting slices wrap up around (but do not intersect) the singularity. For example, the maximal slicing condition $K = 0$ yields an elliptic equation for $\alpha$ that must be solved on each slice. Since the solution of elliptic equations tends to be computationally expensive, hyperbolic, parabolic, and algebraic conditions for $\alpha$ have been developed which also produce singularity avoiding slices. The most popular is the "1 + log" slicing, in which $\alpha = 1 + \ln\gamma$ [40].

Singularity avoiding slices with $\beta^i = 0$ have enabled evolutions of BHs up to $t \sim 30 - 40M$, including a BBH grazing collision [11]. However, as the evolution proceeds the slices become increasingly distorted, leading to large gradients in the metric. Eventually the resulting grid stretching produces numerical errors that cause the code to crash.

A nonzero shift vector can be used to keep grid points from falling towards a BH.

Elliptic, hyperbolic, and parabolic conditions have been developed for $\beta^i$ that counter the grid stretching properties of singularity avoiding slices. Using the hyperbolic $\Gamma$−driver condition for $\beta^i$ along with 1 + log slicing has extended the running time of single BH evolutions to several thousand $M$ [23], and BBH evolutions to several hundred $M$ [25].

## Adaptive Mesh Refinement

Numerical relativity simulations of BBH mergers must typically solve $\sim 17$ or more evolution equations on a 3-D spatial grid. The need to provide adequate resolution of the multiple physical scales in these models implies that some form of adaptivity is required to carry out these calculations, even with today's high performance computers.

For example, the gravitational waves produced typically have wavelengths $\sim 10-100$ times the scale of the BHs themselves. A sufficiently large grid is necessary so that these signals can propagate into the wave zone and be extracted at large distances from the source. The grid must also be large enough so that the boundary conditions at the outer edges of the grid can be applied in an effectively asymptotic region. To accomplish these objectives, higher resolution can be used in the central interaction region, with coarser resolution in the wave zone.

In general, the BHs in a binary will have different masses. For intermediate mass and supermassive BBHs, the component masses can easily differ by factors of $\sim 10-100$ or more, pointing to the need for different spatial resolution in the vicinity of each BH. In the early stages of the merger, the BHs have a relatively wide separation. Ideally, each BH would be surrounded by an appropriately high resolution grid, with lower resolution in the orbital region, and even coarser resolution in the wave zone. The grid resolution would then change adaptively as the merger takes place to accomodate the evolving physical situation.

The use of mesh refinement in 3-D numerical relativity simulations of BHs is still in its early stages. Fixed mesh refinement has been applied to single BH evolution [41] and a short part of a BBH evolution [42]. Adaptive mesh refinement also was used in the propagation of a source-free gravitational wave across a grid [43].

In general, gravitational waves produced in a BBH merger will need to cross one or more mesh refinement boundaries as they propagate from the higher resolution of the interaction region into the lower resolution of the wave zone. Interpolation conditions must be set on the numerical data at the mesh refinement boundaries to allow the waves to cross them smoothly. Linear interpolation conditions, typically used in hydrodynamics codes, result in significant spurious reflections for gravitational waves crossing refinement boundaries. The use of quadratic interpolation conditions at the mesh refinement boundaries dramatically reduces the spurious reflected waves [44].

## Additional Challenges

There are several other areas which are considered critical for the success of numerical relativity BBH simulations. One key area is the role of the constraint equations, and in particular how best to incorporate these relationships into the solution of the Einstein equations [45, 46]. The role of the outer boundary conditions (that is, as $r \to \infty$) and their relationship to preserving the constraint equations during an evolution is also under investigation [47].

While most numerical relativity simulations to date have been carried out using fairly standard finite difference techniques, some groups are applying spectral methods in their codes (e.g. [48]). Overall, since BBH simulations typically require high performance computing resources, the resulting codes must be implemented in a parallel computing environment with attention to speed and the efficient use of memory.

## Towards Astrophysical BBH Mergers

While the use of numerical relativity to model BBH coalescence has proved to be very challenging, significant progress has been made over the past decade. Although important issues remain, currently available techniques are sufficient to begin the study of astrophysically relevant BBH merger waveforms. An important step in this direction was taken by combining full numerical simulations of the merger of two BHs with perturbative techniques to evolve the late-time state of the system. This so-called *Lazarus* approach yielded the first model of a gravitational waveform from an astrophysically plausible merger [12].

This method begins with initial data for a BBH near the plunge. The idea is to use a full 3-D numerical relativity simulation to evolve this system up to the point at which the BHs are close enough to be treated as a single, highly distorted BH. The evolution of this remnant (that is, the ringdown stage) is then carried out using techniques from BH perturbation theory. In this approach, the large scale simulation is reserved for the strongly nonlinear part of the evolution, and simpler numerical techniques are applied to handle the ringdown and extract the gravitational waveform.

The Lazarus approach was first used to evolve the final plunge of 2 identical Schwarzschild BHs. The result of this calculation was that $\sim 3\%$ of the system's total energy and $\sim 12\%$ of its total angular momentum was radiated away in the form of gravitational waves, leaving behind a rotating Kerr BH with rotation parameter $a/M \sim 0.72$ [12]. More recently, these calculations have been extended to include the plunge of BHs with spin [13].

So far, the full numerical relativity evolution used in these calculations lasts only $\sim 13M$, as the original ADM formalism is employed. Since the dynamics of realistic BBHs mergers is expected to take significantly longer than this, these results must be considered preliminary. Nevertheless, they represent an important advance and we can expect to see the Lazarus approach used in longer-lived, more realistic simulations in the future.

# ACKNOWLEDGMENTS

It is a pleasure to thank my colleagues in numerical relativity for many stimulating and informative discussions on the topics presented here. Special thanks go to John Baker, David Brown, Bernd Bruegmann, Dae-Il Choi, Pablo Laguna, Richard Matzner, and Mark Scheel.